%
\input epsf
%
%
%
\def\today{\ifcase\month\or January\or February\or March\or April\or May\or
June\or July\or August\or September\or October\or November\or December\fi
\space\number\day, \number\year}
%
%
\newcount\notenumber

\def\note{\global\advance\notenumber by 1 \footnote{$^{\the\notenumber}$}}
%
%
\def\alphasecnumber{{\rm\ifcase\secnumber\or A\or B\or C\or D\or E\or F\or
G\fi}}
\newif\ifsectionnumbering
\newcount\eqnumber
\def\cleareqnumber{\eqnumber=0}
\def\numbereq{\global\advance\eqnumber by 1
\ifinappendix
  \ifsectionnumbering\eqno(\alphasecnumber.\the\eqnumber)\else
  \eqno(\the\eqnumber)\fi
\else
  \ifsectionnumbering\eqno(\the\secnumber.\the\eqnumber)\else
  \eqno(\the\eqnumber)\fi
\fi}
\def\eqalinno{{\global\advance\eqnumber by 1}
\ifsectionnumbering(\the\secnumber.\the\eqnumber)\else(\the\eqnumber)\fi}
\def\name#1{\ifsectionnumbering\xdef#1{\the\secnumber.\the\eqnumber}
\else\xdef#1{\the\eqnumber}\fi}

\sectionnumberingtrue
%
%
\newcount\refnumber

\immediate\openout1=refs.tex
\immediate\write1{\noexpand\frenchspacing}
\immediate\write1{\parskip=0pt}
\def\ref#1#2{\global\advance\refnumber by 1%
[\the\refnumber]\xdef#1{\the\refnumber}%
\immediate\write1{\noexpand\item{[#1]}#2}}
\def\tie{\noexpand~}

%
%
\font\twelvebf=cmbx10 scaled \magstep1
\newcount\secnumber

\def\newsection#1.{\ifsectionnumbering\cleareqnumber\else\fi%
	\global\advance\secnumber by 1%
	\bigbreak\bigskip\par%
	\line{\twelvebf \the\secnumber. #1.\hfil}\nobreak\medskip\par\noindent}
%
%
%
\def \sqr#1#2{{\vcenter{\vbox{\hrule height.#2pt
	\hbox{\vrule width.#2pt height#1pt \kern#1pt
		\vrule width.#2pt}
		\hrule height.#2pt}}}}
\def\Box{{\mathchoice\sqr54\sqr54\sqr33\sqr23}\,}
%
%
%
\newdimen\fullhsize
\def\fiddle{\fullhsize=6.5truein \hsize=3.2truein}
\def\fullline{\hbox to\fullhsize}
\def\mkhdline{\vbox to 0pt{\vskip-22.5pt
	\fullline{\vbox to8.5pt{}\the\headline}\vss}\nointerlineskip}
\def\mkftline{\baselineskip=24pt\fullline{\the\footline}}
\let\lr=L \newbox\leftcolumn
\def\twocolumns{\fiddle
	\output={\if L\lr \global\setbox\leftcolumn=\columnbox
		\global\let\lr=R \else \doubleformat \global\let\lr=L\fi
		\ifnum\outputpenalty>-20000 \else\dosupereject\fi}}
\def\doubleformat{\shipout\vbox{\mkhdline
		\fullline{\box\leftcolumn\hfil\columnbox}
		\mkftline} \advancepageno}
\def\columnbox{\leftline{\pagebody}}
\magnification=1200
\def\pr#1 {Phys. Rev. {\bf D#1},\tie}
\def\pe#1 {Phys. Rev. {\bf #1\tie}}
\def\pre#1 {Phys. Rep. {\bf #1\tie}}
\def\pl#1 {Phys. Lett. {\bf B#1},\tie}
\def\prl#1 {Phys. Rev. Lett. {\bf #1},\tie}
\def\np#1 {Nucl. Phys. {\bf B#1},\tie}
\def\ap#1 {Ann. Phys. (NY) {\bf #1\tie }}
\def\cmp#1 {Commun. Math. Phys. {\bf #1\tie }}
\def\imp#1 {Int. Jour. Mod. Phys. {\bf A#1\tie }}
\def\mpl#1 {Mod. Phys. Lett. {\bf A#1\tie}}
\def\jhep#1 {JHEP {\bf #1},\tie}
\def\nuo#1 {Nuovo Cimento {\bf B#1\tie}}
\def\class#1 {Class. Quant. Grav. {\bf #1},\tie}
\def\for#1 {Fortsch. Phys. {\bf #1},\tie}
\def\tie{\noexpand~}
\def\ov{\bar}

\parskip=15pt plus 4pt minus 3pt
\headline{\ifnum \pageno>1\it\hfil Linearized Gravity in
the Karch-Randall Braneworld\else \hfil\fi}
\font\title=cmbx10 scaled\magstep1
\font\tit=cmti10 scaled\magstep1
\footline{\ifnum \pageno>1 \hfil \folio \hfil \else
\hfil\fi}
\raggedbottom


\overfullrule0pt

\def\ft#1#2{{\textstyle{{#1}\over{#2}}}}
\def\fft#1#2{{{#1}\over{#2}}}
\def\sgn{{\,\rm sgn}}
\def\sech{{\rm sech}}
\newif\ifinappendix\inappendixfalse
\def\newappendix#1.{\ifinappendix\else\secnumber=0\inappendixtrue\fi%
        \ifsectionnumbering\cleareqnumber\else\fi%
	\global\advance\secnumber by 1%
	\bigbreak\bigskip\par%
	\line{\twelvebf #1.\hfil}\nobreak\medskip\par\noindent}
\def\lref#1#2{\global\advance\refnumber by 1%
\xdef#1{\the\refnumber}%
\immediate\write1{\noexpand\item{[#1]}#2}}
\font\btitle=cmbx12 scaled\magstep1


\rightline{\vbox{\hbox{RU02-08-B}\hbox{MCTP-02-62}\hbox{PUPT-2062}\hbox{hep-th/0211196}}}
\vfill
\centerline{\btitle Linearized Gravity in the Karch-Randall Braneworld}
\vfill
{\centerline{\title Ioannis Giannakis${}^{a}$, James T. Liu${}^{b,c}$
and Hai-cang Ren${}^{a}$ \footnote{$^{\dag}$}
{\rm e-mail: \vtop{\baselineskip12pt
\hbox{giannak@summit.rockefeller.edu, jimliu@umich.edu,}
\hbox{ren@summit.rockefeller.edu}}}}}
\medskip
\centerline{$^{(a)}${\tit Physics Department, The Rockefeller University}}
\centerline{\tit 1230 York Avenue, New York, NY
10021--6399}
\medskip
\centerline{$^{(b)}${\tit Michigan Center for Theoretical Physics}}
\centerline{\tit Randall Laboratory, Department of Physics,
University of Michigan}
\centerline{\tit Ann Arbor, MI 48109--1120} 
\medskip
\centerline{$^{(c)}${\tit Department of Physics, Princeton University,
Princeton, NJ 08544}}
\vfill
\centerline{\title Abstract}
\bigskip
{\narrower\narrower\noindent
We present a linearized gravity investigation of the bent braneworld,
where an AdS$_4$ brane is embedded in AdS$_5$.  While we focus on static
spherically symmetric mass distributions on the brane, much of the
analysis continues to hold for more general configurations.  In addition
to the identification of the massive Karch-Randall graviton and a tower
of Kaluza-Klein gravitons, we find a radion mode that couples to the trace
of the energy-momentum tensor on the brane.  The Karch-Randall radion
arises as a property of the embedding of the brane in the bulk space,
even in the context of a single brane model.
\par}
\vfill\vfill\break


\newsection Introduction.%
In the Randall-Sundrum model
\ref{\rstwo}{L. Randall and R. Sundrum, 
{\sl An alternative to compactification},
\prl83 4690 (1999) [arXiv:hep-th/9906064].
},
gravity is trapped on a (flat) 3-brane embedded in an uncompact
five-dimensional spacetime.  In particular, a linearized treatment of
gravity indicates the presence of a massless graviton localized on the
brane, along with a tower of Kaluza-Klein modes in the bulk
\lref{\garriga}{J. Garriga and T. Tanaka,
{\sl Gravity in the brane-world},
\prl84 2778 (2000) [arXiv:hep-th/9911055].
}%
\lref{\gkr}{S.B. Giddings, E. Katz and L. Randall,
{\sl Linearized gravity in brane backgrounds},
\jhep0003 023 (2000) [arXiv:hep-th/0002091].
}%
\lref{\muck}{W. M\"uck, K.S. Viswanathan and I.V. Volovich,
{\sl Geodesics and Newton's law in brane backgrounds},
\pr62 105019 (2000) [arXiv:hep-th/0002132].
}%
[\rstwo--\muck].
{}From a four-dimensional perspective, the massless graviton yields an
ordinary $1/r$ Newtonian potential, while the Kaluza-Klein modes produce
a $1/r^3$ correction, so that
\lref{\duffliu}{M.J. Duff and J.T. Liu,
{\sl Complementarity of the Maldacena and Randall-Sundrum pictures},
\prl85 2052 (2000) [\class18 3207 (2001)] [arXiv:hep-th/0003237].
}%
[\rstwo,\garriga,\duffliu]
$$
V(r)=\fft{Gm_1m_2}r\left(1+\fft{2L_5^2}{3r^2}\right),
\numbereq\name{\newtcor}
$$
where $L_5$ is the AdS$_5$ ``radius''.

In a complementary approach, corrections to the Newtonian potential may
also be obtained through the investigation of braneworld black holes.
However, an additional benefit of this approach is that the
understanding of black holes would also elucidate the nature of gravity
in the bulk, and not just on the brane.  Furthermore, black holes may
play an important role in both the astrophysics and cosmology of
the braneworld.  Although exact solutions have been found in lower
dimensions
\lref{\ehmone}{R. Emparan, G.T. Horowitz and R.C. Myers,
{\sl Exact description of black holes on branes},
\jhep0001 007 (2000) [arXiv:hep-th/9911043].
}%
\lref{\ehmtwo}{R. Emparan, G.T. Horowitz and R.C. Myers,
{\sl Exact description of black holes on branes. II: Comparison with BTZ
black holes and black strings},
\jhep0001 021 (2000) [arXiv:hep-th/9912135].
}%
[\ehmone,\ehmtwo],
none are as yet known for black holes on 3-branes.  Nevertheless, much
information has been gleaned from appropriate limits
\lref{\chr}{A. Chamblin, S.W. Hawking and H.S. Reall,
{\sl Brane-world black holes},
\pr61 065007 (2000) [arXiv:hep-th/9909205].
}%
\lref{\ighcr}{I. Giannakis and H.-C. Ren,
{\sl Recovery of the Schwarzschild metric in theories with localized
gravity beyond linear order},
\pr63 024001 (2000) [arXiv:hep-th/0007053].
}%
\lref{\gren}{I. Giannakis and H.-C. Ren,
{\sl Possible extensions of the 4-D Schwarzschild Horizon in the
Brane World}, \pr63 125017 (2001) [arXiv:hep-th/0010183].
}%
\lref{\tanaka}{H. Kudoh and T. Tanaka,
{\sl Second Order Perturbations in the Randall-Sundrum Infinite
Brane Model}, \pr64 084022 (2001) [arXiv:hep-th/0104049].
}%
\lref{\bgm}{M. Bruni, C. Germani and R. Maartens,
{\sl Gravitational collapse on the brane},
\prl87 231302 (2001) [arXiv:gr-qc/0108013].
}%
\lref{\deruelle}{N. Deruelle,
{\sl Stars on branes: The view from the brane},
arXiv:gr-qc/0111065.
}%
\lref{\kt}{P. Kanti and K. Tamvakis,
{\sl Quest for localized 4-D black holes in brane worlds},
\pr65 084010 (2002) [arXiv:hep-th/0110298].
}%
\lref{\cfm}{ R. Casadio, A. Fabbri and L. Mazzacurati,
{\sl New black holes in the brane-world?},
\pr65 084040 (2002) [arXiv:gr-qc/0111072].
}%
\lref{\wis}{T. Wiseman,
{\sl Relativistic stars in Randall-Sundrum gravity},
\pr65 124007 (2002) [arXiv:hep-th/0111057].
}%
[\chr--\wis].
In particular, Ref.~[\ighcr] shows that the Schwarzschild metric
is recovered on the brane to first nonlinear order, thus demonstrating
the consistency of the braneworld with classical tests of general
relativity.

In the Randall-Sundrum model, flatness of the 3-brane is maintained
by fine-tuning the Randall-Sundrum brane tension.  Subsequently, it was
realized that ``bent-braneworlds'' (namely 3-branes with cosmological
constant) may be obtained by relaxing the fine-tuning condition
\lref{\nihei}{T. Nihei,
{\sl Inflation in the five-dimensional universe with an orbifold extra
dimension},
\pl465 81 (1999) [arXiv:hep-ph/9905487].
}%
\lref{\kaloper}{N. Kaloper,
{\sl Bent domain walls as braneworlds},
\pr60 123506 (1999)\hfil\noexpand\break
[arXiv:hep-th/9905210].
}%
\lref{\kimkim}{H.B. Kim and H.D. Kim,
{\sl Inflation and gauge hierarchy in Randall-Sundrum compactification},
\pr61 064003 (2000) [arXiv:hep-th/9909053].
}%
\lref{\dewolfe}{O. DeWolfe, D.Z. Freedman, S.S. Gubser and A. Karch,
{\sl Modeling the fifth dimension with scalars and gravity},
\pr62 046008 (2000) [arXiv:hep-th/9909134].
}%
\lref{\karch}{A. Karch and L. Randall,
{\sl Locally localized gravity},
\jhep0105 008 (2001) [Int. J. Mod. Phys. {\bf A16}, 780 (2001)]
[arXiv:hep-th/0011156].
}%
[\nihei--\karch].
In Ref.~[\karch], Karch and Randall investigated the nature of gravity
in this class of bent braneworlds.  They demonstrated that for both flat
and de~Sitter 3-branes embedded in five-dimensional AdS, there is a
single trapped massless graviton.  However, for an AdS 3-brane embedded
in AdS$_5$ (which is referred to as the AdS$_{\rm AdS}$ braneworld), a
novel feature occurs --- while a graviton is indeed bound to the brane, it
turns out that the trapped graviton is {\it massive}, with mass [\karch]
$$
M^2\approx\fft{3L_5^2}{2L_4^4}.
\numbereq\name{\krmasse}
$$
This mass for the graviton has subsequently been obtained from a
holographic interpretation of the Karch-Randall model, both from the
bulk
\lref{\porhol}{M. Porrati,
{\sl Mass and gauge invariance. IV: Holography for the Karch-Randall
model},
\pr65 044015 (2002) [arXiv:hep-th/0109017].
}%
\lref{\bora}{R. Bousso and L. Randall,
{\sl Holographic domains of anti-de Sitter space},
JHEP {\bf 0204}, 057 (2002) [arXiv:hep-th/0112080].
}%
[\porhol,\bora]
and the dual theory on the brane
\lref{\porhiggs}{M. Porrati,
{\sl Higgs phenomenon for 4-D gravity in anti de Sitter space},
\jhep0204 058 (2002) [arXiv:hep-th/0112166].
}%
\lref{\duffliutwo}{M.J. Duff and J.T. Liu,
{\sl Complementarity of the Maldacena and Karch-Randall pictures},
arXiv:hep-th/0207003.
}%
[\porhiggs,\duffliutwo].
For an alternative derivation see also
\ref{\mie}{A. Miemiec,
{\sl A power law for the lowest eigenvalue in localized massive
gravity},
\for49 747 (2000) [arXiv:hep-th/0011160].
}.
The Newtonian law for a dS$_4$ brane embedded in a five-dimensional spacetime
was investigated in
\lref{\keha}{A. Kehagias and K. Tamvakis,
{\sl Graviton localization and Newton law for a dS$_4$ brane in 5D bulk},
\class19 L185 (2002) [arXiv:hep-th/0205009].
}%
\lref{\ito}{M. Ito,
{\sl Localized Gravity on de Sitter Brane in Five Dimensions},
arXiv:hep-th/0204113.
}%
[\keha,\ito].
Although theories with massive gravitons potentially suffer a
van~Dam-Veltman-Zakharov discontinuity
\lref{\vdv}{H. van Dam and M. Veltman,
{\sl Massive And Massless Yang-Mills And Gravitational Fields},
\np22 397 (1970).
}%
\lref{\zak}{V.I. Zakharov,
{\sl Linearized gravitation theory and the graviton mass},
JETP Lett. {\bf 12}, 312 (1970).
}%
[\vdv,\zak],
it was shown that this discontinuity is absent in the presence of a
non-vanishing cosmological constant
\lref{\kogan}{I.I. Kogan, S. Mouslopoulos and A. Papazoglou,
{\sl The m $\to$ 0 limit for massive graviton in dS$_4$ and AdS$_4$: How to
circumvent the van Dam-Veltman-Zakharov discontinuity},
\pl503 173 (2001) [arXiv:hep-th/0011138].
}%
\lref{\porvdv}{M. Porrati,
{\sl No van Dam-Veltman-Zakharov discontinuity in AdS space},
\pl498 92 (2001) [arXiv:hep-th/0011152].
}%
[\kogan,\porvdv].
While a graviton mass arising from explicit breaking of general
covariance leads to inconsistencies at the quantum level
\ref{\ddls}{F.A. Dilkes, M.J. Duff, J.T. Liu and H. Sati,
{\sl Quantum discontinuity between zero and infinitesimal graviton mass
with  a Lambda term},
\prl87 041301 (2001) [arXiv:hep-th/0102093].
},
no such difficulties arise when the mass is generated dynamically, as in
the holographic Karch-Randall model [\porhiggs,\duffliutwo].
Nevertheless, many issues still arise in understanding models of massive
gravity.  

In this paper, we shall explore linearized gravity in the Karch-Randall
model, and in particular we shall focus on the recovery of the
Schwarzschild-AdS$_4$ geometry on the brane.  In order to pursue this
solution, we consider static radially-symmetric four-dimensional
configurations, both in vacuum and with a point mass source on the
brane.  Our analysis of the off-brane profile of the gravitational field
indicates the existence of a physical radion in the
Karch-Randall model. Similar results were obtained by
Chacko and Fox
\ref{\chacko}{Z. Chacko and P. Fox,
{\sl Wave Function of the Radion in the dS and AdS Brane Worlds},
\pr64 024015 (2001) [arXiv:hep-th/0102023].
},
in the limit of infinite separation of two AdS$_4$
branes embedded in AdS$_5$ (see also
\ref{\defa}{P. Binetruy, C. Deffayet
and D. Langlois,
{\sl The Radion in Brane Cosmology},
\np615 219 (2001) [arXiv:hep-th/0101234].
}).
While this mode was regarded as
a gauge artifact in Ref.~[\karch], we show here that it cannot be completely
gauged away, even when considering the effects of brane bending.  We
demonstrate that the effect of the radion shows up as a correction to
the Schwarzschild-AdS$_4$ solution of comparable strength as from the
standard tower of Kaluza-Klein modes. 

We begin in section 2 by providing our bent braneworld conventions and
setting up the general radially-symmetric metric ansatz.  Within this
ansatz, we obtain equations governing the Kaluza-Klein modes, as well
as the radial wavefunctions on the brane.  While much of the content
here is standard, this provides the context on which the rest of the
paper is based.  In sections 3, we investigate the quasi-zero mode
graviton, and in section 4 the Kaluza-Klein modes.  Then in section 5
we demonstrate the presence of the Karch-Randall radion in the spectrum
of transverse-traceless fluctuations.  In section 6, we combine the
graviton, radion and KK modes and obtain their coupling to a matter
(point) source on the brane.  Finally, we conclude in section 7.

\newsection Einstein Equations, Symmetries and Boundary Conditions.%
The starting point for the Randall-Sundrum class of braneworlds is a
warped product metric of the form
$$
ds^2=e^{2A(y)}{\bar g}_{\mu\nu}(x,y)dx^\mu dx^\nu+dy^2,
\numbereq\name{\eqena}
$$
which is the most general metric with four-dimensional covariance.
Here and throughout the
paper, we adopt the convention that the Greek indices take values
0--3 and the Latin indices 0--4.  Thus $g_{mn}$ is the full
five-dimensional metric, while $\bar g_{\mu\nu}$ is the four-dimensional
metric ``on the brane''.

For the case where $\bar g_{\mu\nu}(x)$ is independent of $y$, the
five-dimensional Ricci tensor is simply expressed in terms of the four
dimensional metric ${\bar g}_{\mu\nu}$ and warp factor $A(y)$ as
$$
\eqalign{
R_{\mu\nu}&={\bar R_{\mu\nu}}-g_{\mu\nu}(4\dot A^2+\ddot A),\cr
R_{44}&=-4(\dot A^2+\ddot A),}
\numbereq\name{\eqfiond}
$$
where a dot indicates differentiation with respect to $y$.
Away from the brane, the space is simply AdS$_5$, so that the
five dimensional metric $g_{mn}$ satisfies the Einstein condition
$$
R_{mn}=\Lambda_5 g_{mn},
\numbereq\name{\eqdyo}
$$
where $\Lambda_5$ is the bulk cosmological constant.  Furthermore, for a
bent braneworld, we assume the four dimensional metric $\bar g_{\mu\nu}$
similarly satisfies an Einstein condition
$$
{\bar R_{\mu\nu}}={\Lambda}_4{\bar g}_{\mu\nu},
\numbereq\name{\eqassis}
$$
this time with brane cosmological constant $\Lambda_4$.  As a result,
the bulk Einstein equations take on the form
$$
\eqalign{
\Lambda_4e^{-2A}-4\dot A^2-\ddot A&=\Lambda_5,\cr
-4\dot A^2-4\ddot A &=\Lambda_5.}
\numbereq
$$
Eliminating the second derivative (and hence removing sensitivity to
any possible jump discontinuity in the warp factor), we obtain a simple
equation
$$
f^2-\Bigl({\dot f\over \kappa}\Bigr)^2 = \Bigl({k\over \kappa}\Bigr)^2,
\numbereq\name{\wfeqn}
$$
where $f=e^A$, and we have parameterized the bulk and brane cosmological
constants in terms of the respective AdS radii%
\note{Here we have assumed an AdS brane embedded in an AdS bulk,
since this is what gives rise to a massive graviton.  We will not
elaborate on dS or flat braneworlds, but simply note that such solutions
are obvious from Eq.~(\wfeqn) with slight modification of signs.}
$$
\Lambda_5=-4\kappa^2=-\fft4{L_5^2},\qquad\Lambda_4=-3k^2=-\fft3{L_4^2}.
\numbereq
$$
Eq.~(\wfeqn) yields a simple solution for the warp factor
$$
e^{2A} = \left({k\over \kappa}\right)^2{\cosh^2{\kappa(|y|-y_0)}},
\numbereq\name{\eqdisitter}
$$
where $y_0$ is a constant related to the value of the warp factor on the
brane (which is located at $y=0$).  A natural choice is to demand
$e^{2A}=1$ on the brane, so that $\cosh{\kappa y_0} = \kappa/k$.  In the
nearly-flat limit, this may be approximated as
$$
e^{-2\kappa y_0}\approx\left(\fft{k}{2\kappa}\right)^2\qquad
(k/\kappa\ll1).
\numbereq\name{\kkyrel}
$$
Note that we have inserted an
absolute value in the $y$ coordinate, which enforces the location of the
brane.  This discontinuity in the slope of the warp factor may be fixed
by the Israel matching conditions accounting for the brane tension.

We treat the above solution as the braneworld vacuum, and consider
the effect of an axially symmetric perturbation about this
background.  First note that the AdS$_4$ metric, $\bar g_{\mu\nu}$,
may be written in global coordinates as
$$
\bar d\bar s^2=-(1+k^2r^2)dt^2+{dr^2\over1+k^2r^2}+r^2d\Omega^2.
\numbereq\name{\eqglobads}
$$
With this parametrization in mind, we take the most general axially
symmetric and static metric in $D=4+1$ with the following form
$$
ds^2=e^{2A}[-e^{a+{\bar a}}dt^2
+e^{b-{\bar a}}dr^2+e^{c}r^2d\Omega_2^2]+dy^2,
\numbereq\name{\eqpente}
$$
where $e^{2A}$, given by Eq.~(\eqdisitter), and
${\bar a}={\ln({1+k^2r^2})}$ are fixed background quantities.
This general ansatz depends on three
functions $a$, $b$ and $c$, all functions of $r$ and $y$.

Substituting the metric (\eqpente) into the Einstein equation,
(\eqdyo), and keeping terms linear in $a$, $b$ and $c$, we obtain
the following equations
$$
\eqalign{
&e^{\bar a}\Big[a^{\prime\prime}
+{2\over r}a^\prime +{1\over 2}{\bar a}^\prime(3a'-b'+2c')\Big]
+e^{2A}\Big[\ddot a+ \dot A(5\dot a+\dot b+2\dot c)\Big]\cr
&\kern3.5truein
=6k^2b+16\pi G_5e^{2A}(-T^t_t+\ft13T)\delta(y),\cr
&e^{\bar a}\Bigl[a^{\prime\prime}+2c^{\prime\prime}-{2\over r}(b'-2c')
+{1\over 2}{\bar a}^\prime(3a^\prime-b^\prime+2c^\prime)\Bigr]
+e^{2A}\Bigl[\ddot b+\dot A(\dot a+5\dot b+2\dot c)\Bigr]\cr
&\kern3.5truein
=6k^2 b+16\pi G_5e^{2A}(-T^r_r+\ft13T)\delta(y),\cr
&e^{\bar a}\Bigl[c^{\prime\prime}+{1\over r}(a'-b'+4c')
+{\bar a}^{\prime}c^\prime \Bigr]+
e^{2A}\Bigl[\ddot c+\dot A(\dot a+\dot b+6\dot c) \Bigr]
-{2\over r^2}(b-c)\cr
&\kern3.5truein
=6k^2b+16\pi G_5e^{2A}(-T^\theta_\theta+\ft13T)\delta(y),\cr
&\ddot a+\ddot b+2\ddot c+2\dot A(\dot a
+\dot b+2\dot c)=16\pi G_5(\ft13T)\delta(y),\cr
&\dot a^\prime+2\dot c^\prime-{2\over r}(\dot b-\dot c)
+{1\over 2}{\bar a}^\prime(\dot a-\dot b)=0,\cr}
\numbereq\name{\eqenea}
$$
where primes denote $r$ derivatives.  Note that we have included a brane
stress energy tensor, $T^\mu_\nu\delta(y)$, as a source.  This is in
addition to the Karch-Randall brane tension itself, which is accounted
for by the kink in the warp factor, (\eqdisitter).  By integrating the
equations (\eqenea) across the brane, one would obtain jump conditions
on the metric functions $a$, $b$ and $c$.  Such conditions, in fact, are
equivalent to the Israel matching conditions arising from the matter
on the brane.  These equations will be the starting point of our
investigation.

We now discuss the gauge symmetries of these equations---more
specifically the coordinate transformations that respect the
axially symmetric, static form of the metric.  In general, consider
a coordinate transformation (gauge transformation) generated
by functions $v$ and $u$, such that
$$
\eqalign{
r&\to r+v(r, y),\cr
y&\to y+u(r, y).}
\numbereq
$$
By demanding that this particular coordinate transformation respects the
form of the five dimensional metric, (\eqpente), we find that the functions
$v$ and $u$ necessarily obey the following relations to linear order
$$
e^{2A-{\bar a}}{\dot v}+u^\prime=0, \qquad
{\dot u}=0.
\numbereq\name{\eqreann}
$$
Furthermore, the components of the metric transform under these
residual coordinate transformations as follows
$$
\eqalign{
&a(r, y) \to a(r, y)+{\bar a}^\prime{v}+2{\dot A}u,\cr
&b(r, y) \to b(r, y)-{\bar a}^{\prime}v+2{\dot A}u+2v^\prime,\cr
&c(r, y) \to c(r, y)+2{\dot A}u+{2\over r}v.\cr}
\numbereq\name{\eqrbcv}
$$
The constraints, (\eqreann), may be integrated to provide the following
form for the parameters of transformation
$$
u=\chi(r), \qquad v(r, y)=-e^{\bar a}{\chi}^{\prime}{\int^y}d{x}\,
e^{-2A(x)}+{\phi}(r).
\numbereq\name{\eqriot}
$$
Thus axially symmetric gauge transformations may be fully parameterized
by two functions of $r$, namely $\chi(r)$ and $\phi(r)$.  We will make
use of such gauge transformations below, when working out the brane
bending mode [\garriga,\gkr].

We now proceed to find solutions to the linearized
equations of motion. We begin with the penultimate equation of (\eqenea),
corresponding to $R_{yy}$, which may be integrated to give
$$
a+b+2c={8\pi G_5T(r)\over3k^2}{\dot A}(y)\sgn(y)+f(r).
\numbereq\name{\eqvirile}
$$ 
To arrive at this expression, we had to make use of the explicit form
(\eqdisitter) for $A(y)$.  Here
$f(r)$ is an arbitrary function and $T(r)$ is the trace of the
energy-momentum tensor on the brane.
Since $\dot A(y)\sgn(y)\to1$ as $y\to\infty$, we may satisfy the
asymptotic conditions
$$
\lim_{|y|, r\to\infty}a=\lim_{|y|, r\to\infty}b
=\lim_{|y|, r\to\infty}c=0,
\numbereq\name{\eqexy}
$$
provided $f(r)=-8\pi G_5T(r)/3k^2$. Thus we may always
set $a+b+2c\approx0$ away from the brane.  However, in the presence of
matter on the brane, the trace of the energy-momentum tensor provides
an obstruction, and this will no longer be possible except in the
asymptotic region.

We will return to this point when considering brane bending in the
presence of sources on the brane.  For now, we choose to examine
radially symmetric excitations of the vacuum.  Thus we set $T(r)=0$,
and in turn are allowed to simply choose $a+b+2c=0$.  We refer to
this gauge choice as the traceless condition.  By substituting
$c=-{1\over 2}(a+b)$ into the final equation of (\eqenea) and
integrating, we find
$$
b^\prime+{1\over r}(a+3b)-{1\over 2}{\bar a}^\prime(a-b)=0.
\numbereq\name{\eqrichos}
$$
Furthermore, by taking into account that ${\bar a}=
{\ln(1+k^2r^2)}$, we may eliminate $a$ in terms of $b$
$$
a=b-(1+k^2r^2)(rb^\prime+4b).
\numbereq\name{\eqriotx}
$$
This allows us to rewrite the equations of motion in terms of the single
function $b(r,y)$.  We now separate variables for the
transverse and traceless modes by decomposing
$b(r, y)={\sum_\mu}B(\mu)b(r|\mu){\psi}(y|\mu)$.
Separating variables on the second equation of (\eqenea) then results in
a second order differential equation for the modes $b(r|\mu)$
$$
b^{\prime\prime}+{4\over r}{{1+2k^2r^2}\over {1+k^2r^2}}
b^\prime
+{{10k^2-{\mu}^2}\over {1+k^2r^2}}b=0,
\numbereq\name{\eqvios}
$$
as well as an eigenvalue equation for the modes ${\psi}(y|\mu)$
$$
e^{2A(y)}({\ddot{\psi}}+4{\dot A}{\dot{\psi}})=
-{\mu}^2{\psi}.
\numbereq\name{\eqwilde}
$$
{}From a four-dimensional point of view, $\psi(y|\mu)$ are the
wavefunctions of the Kaluza-Klein gravitons, with Kaluza-Klein masses
$\mu$.  Note that the wave equation on the brane, (\eqvios), may be
written in the form
$$
(\Box_4+10k^2-\mu^2)b+\fft2r(1+2k^2r^2)b'=0,
\numbereq
$$
where $\Box_4$ is the {\it scalar} Laplacian on the AdS$_4$ background.
This is effectively the explicit form of the Lichnerowicz operator in
our choice of coordinates with a spherically symmetric background.

Eq.~(\eqwilde) defines a Sturm-Liouville problem with $e^{2A}$ 
as the measure of the normalization integral. Imposing the boundary 
conditions $e^{2A}\psi\to 0$ as $y\to\infty$ and $\dot\psi=0$ at 
$y=0$, we have a normalization integral
$$\int_0^\infty dy\,e^{2A(y)}\psi^2(y|\mu)=\psi(0|\mu){\partial\dot
\psi(0|\mu)\over\partial\mu^2}.
\numbereq\name{\eqconst}
$$
In the following sections, we will examine the eigenvalue
equation (\eqwilde) for the quasi-zero mode, the radion and the
Kaluza-Klein modes, as well as the scalar
equation on the brane, (\eqvios).

\newsection The Karch-Randall Quasi-Zero Mode.%
It was demonstrated in Ref.~[\karch] that the graviton spectrum in the
AdS braneworld is composed of a quasi-zero mode graviton trapped on the
brane as well as a discrete tower of Kaluza-Klein modes.  This spectrum
may be determined by examination of the eigenvalue equation (\eqwilde)
for the graviton wavefunction.  Here we shall examine the quasi-zero
mode in detail and reproduce the expression, (\krmasse), for the
graviton mass
\lref{\mdsch}{M.D. Schwartz,
{\sl The emergence of localized gravity},
\pl502 223 (2001) [arXiv:hep-th/0011177].
}%
[\karch,\porhol,\mdsch],
by elementary means.  The method that we employ does not depend on the
details of the warp factor $A(y)$.  The Kaluza-Klein modes will be
considered in the following section.  Note that we suppress the parameter
$\mu$ whenever no possibility for confusion may arise.

We begin by assuming that the mode equation, (\eqwilde), admits a normalized
solution with eigenvalue $\mu^2 \ll m^2$.  Here, $m^2$ is a scale introduced
by the curvature of the brane [and may be taken to be $k^2$ for the
Karch-Randall warp factor, (\eqdisitter)].  In this case, we can treat
the right hand side of the equation perturbatively.
To zeroth order, the right hand side of (\eqwilde) is absent, and
the equation possesses two solutions
$$
u_1=1, \qquad u_2={\int^y}dx\,e^{-4A(x)},
\numbereq\name{\eqhesse}
$$
with Wronskian $W(u_1, u_2)=e^{-4A}$.
On writing the normalizable combination of the zeroth order
solution as $\psi^{(0)}(y)$, the first order correction
to the solution, $\psi^{(1)}(y)$, is determined by the inhomogeneous equation
$$
{\ddot\psi^{(1)}}+4{\dot A}{\dot\psi^{(1)}}
=-\mu^2e^{-2A}\psi^{(0)}.
\numbereq\name{\eqboulez}
$$
This equation may be solved by the method of variation of coefficients;
the solution is given by
$\psi^{(1)}=\mu^2(c_1u_1+c_2u_2)$, where $c_1$ and $c_2$ are both
functions of $y$ and are given by the expressions
$$
\eqalign{
c_1(y)&=\int^y dx\,{e^{-2A}\psi^{(0)}u_2\over W(u_1,u_2)}=\int^y
dx\,e^{2A}\psi^{(0)}u_2, \cr
c_2(y)&=-\int^y dx\,{e^{-2A}\psi^{(0)}u_1\over W(u_1,u_2)}=-\int^y
dx\,e^{2A}\psi^{(0)}u_1. \cr}
\numbereq\name{\eqscion}
$$
with the constants pertaining the indefinite integrals fixed
by the normalizability condition. The eigenvalue $\mu^2$
is determined by the Israel matching condition of the approximate
solution $\psi=\psi^{(0)}+\psi^{(1)}$
on the brane, which for vanishing matter stress-energy reads simply
$$
0=\dot\psi(0)=\dot\psi^{(0)}(0)+\mu^2c_2(0)\dot u_2(0).
\numbereq\name{\eqxios}
$$
This gives rise to the quasi-zero mode graviton mass
$$
\mu^2=-{\dot\psi^{(0)}(0)\over c_2(0)\dot u_2(0)},
\numbereq\name{\eqsamos}
$$
which is valid provided $\mu^2\ll m^2$.

In the case of the Karch-Randall brane model, the warp factor $e^{2A}$
is given explicitly by (\eqdisitter). The zeroth order solutions to the mode
equation, (\eqwilde), reads simply
$$
u_1=1, \qquad u_2={\tanh{\eta}}\,(2+{\sech^2{\eta}}),
\numbereq\name{\eqhakan}
$$
where $\eta\equiv \kappa(y-y_0)$. As a result, the normalizable zeroth
order solution is given by
$$
\psi^{(0)}=1-\ft12{\tanh{\eta}}\,(2+{\sech^2{\eta}}).
\numbereq\name{\eqzeroth}
$$
Proceeding to the first order correction, we compute
$c_1$ and $c_2$, which are given by the integrals
$$
c_1=-{1\over 3k^2}{\int_{\eta}^\infty }d\zeta\,
{\psi^{(0)}}(\zeta)(2{\cosh^{2}{\zeta}}+1){\tanh{\zeta}},
\qquad c_2={1\over 3k^2}\int_{\eta}^\infty d{\zeta}\,
{\psi}^{(0)}(\zeta){\cosh^{2}{\zeta}}.
\numbereq\name{\eqrioz}
$$

Carrying out the integrations, we find the quasi-zero mode wavefunction
is given to first order in ${\mu^2\over \kappa^2}$ as
$$
\eqalign{
\psi&={\psi^{(0)}}+\mu^2(c_1u_1+c_2u_2)\cr
&=1-\ft12{\tanh{\eta}}\,(2+{\sech^2{\eta}})\cr
&\quad+{\mu^2\over 3k^2}
\Bigl[\ft12\sech^2\eta-2(1-\tanh^2\eta)+\ft16\left(1-\ft12\tanh\eta\,
(2+\sech^2\eta)\right)\cr
&\qquad\qquad+\ln(1+e^{-2\eta})\left(1+\ft12\tanh\eta(2+\sech^2\eta)\right)
\Bigr].\cr}
\numbereq\name{\eqroustou}
$$
The four-dimensional mass parameter $\mu$ is calculated according to
(\eqsamos); we find
$$
\mu^2 \approx 6k^2e^{-2{\kappa}{y_0}} = {3\over 2}{k^4\over \kappa^2},
\numbereq\name{\eqvint}
$$
provided $\mu^2/k^2\ll1$.  The final relation was obtained
using the relation (\kkyrel), and reproduces the expected value of the
Karch-Randall graviton mass, (\krmasse).

Having identified the quasi-zero mode wavefunction, (\eqroustou), and
the graviton mass, (\eqvint), we now proceed to determine the profile of
$b(r)$ on the brane.  This is accomplished by finding solutions of the
AdS$_4$ wave equation, (\eqvios), consistent with boundary conditions.
By introducing a new variable $\xi=-k^2r^2$, and defining
$b=f/(1-\xi)$, Eq.~(\eqvios) may be transformed to the
form
$$
{{\partial^2}\over {{\partial{\xi^2}}}}f+{5\over {2\xi}}
{{\partial}\over {\partial\xi}}f=-{{\ov\epsilon}
\over {{\xi}(1-\xi)}}f,
\numbereq\name{\eqpertu}
$$
where $\ov\epsilon=\mu^2/4k^2$.  While this is easily solved in terms of
hypergeometric functions, for the quasi-zero mode, $\ov\epsilon \ll 1$,
we find it instructive to examine the series solution using the same
perturbative approach developed above.

To zeroth order, there are simply two linearly independent solutions
$$
u_1=1, \qquad u_2=(-\xi)^{-{3\over 2}}.
\numbereq\name{\eqcrios}
$$
Imposing boundary conditions at spatial infinity, we demand for the
zeroth order solution that $b^{(0)}\to0$ as $\xi\to-\infty$.  This
ensures that the spacetime remains asymptotically AdS$_4$ on the brane.
This requirement selects the second solution, so that we take
$$
f^{(0)}=C(-\xi)^{-{3\over 2}}=\fft{C}{k^3r^3},
\numbereq
$$
where $C$ is a constant.
The next order solution can be written as
$$
f^{(1)}=-\fft{4\ov\epsilon C}3\left[(-\xi)^{-1/2}-\tan^{-1}(-\xi)^{-1/2}
+\ft12(-\xi)^{-3/2}\ln(1-\xi)\right].
\numbereq
$$
As a result, $b(r)$ is given up to first order in
$\ov\epsilon\approx3k^2/8\kappa^2$ by
$$
b(r)={C\over {1+k^2r^2}}\left[{1\over
{k^3r^3}}\left(1-\fft{k^2}{4\kappa^2}\ln(1+k^2r^2)\right)
-\fft{k^2}{2\kappa^2}\left({1\over {kr}}-\tan^{-1}\fft1{kr}\right)
\right].
\numbereq\name{\eqpoil}
$$
The remaining functions, $a(r)$ and $c(r)$ may now be determined through
Eq.~(\eqriotx) and the relation $c=-{1\over 2}(a+b)$.  The result is
$$
\eqalign{
a(r)&={C\over {1+k^2r^2}}\left[{1\over {kr}}\left(1-{{k^2}\over
4{\kappa}^2}
\ln(1+k^2r^2)\right)+{{k^2}\over {2\kappa^2}}
(3+2k^2r^2)\left(\fft1{kr}-\tan^{-1}{1\over{kr}}\right)\right],\cr
c(r)&=-{C\over 2}\left[{1\over {k^3r^3}}\left(1-{k^2\over 4\kappa^2}
\ln(1+k^2r^2)\right)+{{k^2}\over {\kappa^2}}\left(
\fft1{kr}-{\tan^{-1}{1\over {kr}}}\right)\right].\cr}
\numbereq\name{\eqcuio}
$$

Given the quasi-zero mode graviton, this above solution ought to
reproduce the Schwarzschild-AdS black hole at linearized
order.  However, in order to compare this metric, computed in the
transverse-traceless gauge, with that in the standard Schwarzschild-AdS
form, we must transform the metric component $c$ to zero.  This may be
accomplished by making use of the residual coordinate
transformations, (\eqrbcv) and (\eqriot), with parameter $\phi(r)$
$$
a \to a+{{2k^2r}\over {1+k^2r^2}}\phi(r), \qquad
b \to b-{{2k^2r}\over {1+k^2r^2}}\phi(r)+2{\phi}^{\prime}(r),
\qquad c \to c+{2\over r}{\phi}(r).
\numbereq\name{\eqdiop}
$$
We see that $c$ may be eliminated by choosing $\phi(r)=-\fft{r}2c(r)$,
whereupon the transformed metric components take the form
$$
\eqalign{
a(r)&=-{{2G_4M}\over{r{(1+k^2r^2)}}}\left[1
-{{k^2}\over 4\kappa^2}\ln(1+k^2r^2)+{k^3r\over \kappa^2}
(1+k^2r^2)\left(\fft1{kr}-\tan^{-1}{1\over kr}\right)\right],\cr
b(r)&={{2G_4M}\over{r{(1+k^2r^2)}}}\left[1+{k^2\over 2\kappa^2}
-{k^2\over 4\kappa^2}\ln(1+k^2r^2)\right].\cr}
\numbereq\name{\eqpoilas}
$$
Here we have identified the dimensionless constant $C$ with the mass
according to $C=-4G_4Mk/3$.  This ensures that, up to terms of
${\cal O}(k^2/\kappa^2)$, this solution reproduces the linearized portion
of Schwarzschild-AdS
$$
ds^2=-\left(1-\fft{2G_4M}r+k^2r^2\right)dt^2
+\left(1-\fft{2G_4M}r+k^2r^2\right)^{-1}dr^2+r^2d\Omega^2.
\numbereq
$$
The normalization of $C$ will be obtained in section 6, when we consider
an explicit source on the brane.

This demonstrates that the quasi-zero mode graviton is responsible for
the long-range gravitational interaction on the brane, even though it
has mass $\mu^2\sim k^4/\kappa^2$.  This is a feature of the AdS$_4$
geometry, in that, for the limit we are considering, the mass is
infinitesimal compared to the natural scale of the AdS curvature
({\it i.e.}~$\ov\epsilon\ll1$ for the dimensionless mass $\ov\epsilon$).
Because of this, the background curvature provides a cutoff to the space
before any effects of the graviton mass may be discerned.  Furthermore,
we see that (at least to this order) no van Dam-Veltman-Zakharov
discontinuity [\vdv,\zak] arises, which is consistent with the results
of Refs.~[\kogan,\porvdv].

\newsection Kaluza-Klein Modes.%
Having examined the quasi-zero mode in detail, we now turn to the
massive Kaluza-Klein spectrum.  Although a perturbative investigation
is no longer valid in this regime, we find that the wave equation for
the Kaluza-Klein modes, (\eqwilde), admits solutions of hypergeometric
form.  To see this, we rewrite the wavefunction $\psi(y)$ according to
$\psi(y)=\phi(y)\sech^2{\kappa(y-y_0)}$ and introduce the variable
$x=\tanh \kappa(y-y_0)$.  As a result, the mode equation, (\eqwilde), is
converted to an associated Legendre equation
$$
(1-x^2){d^2\phi\over dx^2}-2x{d\phi\over dx}+\Big[l(l+1)-{m^2\over
1-x^2}\Big]\phi=0,
\numbereq\name{\eqeric}
$$
with $m=2$ and $l=E_0-2$.  Here, $E_0$ is the lowest energy eigenvalue
for the massive graviton representation $D(E_0,s=2)$ of the AdS$_4$
isometry group SO(2,3), and is given by
$$
E_0=\fft32+{1\over 2}\sqrt{9+{4\mu^2\over k^2}}.
\numbereq\name{\eqmathra}
$$
The range of the new variable $x$ is $-({1-{k^2/ \kappa^2}})^{1/2}<x<1$,
with the AdS$_5$ boundary located at $x=1$ and the brane located at
$x=-(1-{k^2/\kappa^2})^{1/2}\equiv\bar x$. The normalization integral
of $\psi(y)$ over $y$, (\eqconst), becomes that of $\phi(x)$ over $x$
with a constant measure.

The general solution to the associated Legendre equation, (\eqeric),
is a linear combination of $P_l^m(x)$ and $Q_l^m(x)$.  Only $P_l^m(x)$
is retained since the other solution is singular at $x=1$, which causes
the normalization integral to diverge. Therefore the Kaluza-Klein
graviton wavefunctions are of the form
$$
\psi(x)=C(1-x^2)P_l^m(x),
\numbereq\name{\eqbravo}
$$
where the normalization constant may be determined by (\eqconst).
The quantization condition on $l$ arises from the Israel matching
condition on the brane, which is simply ${d\psi/dx}\vert_{x=\bar x}=0$
in the absence of matter on the brane.  We now consider two cases, the
first in the formal absence of a brane (vanishing brane tension), and
the second with a brane.

{\bf \noindent\the\secnumber.1 Without a brane}
\par\noindent
At zero tension, $y_0=0$, and the AdS$_4$ curvature is identified with
that of AdS$_5$, $k=\kappa$.  This provides an interval $0<x<1$ to the
right of the brane.  However, for $y_0=0$, the kink at the brane is
absent, and we may drop the absolute value on $y$ in the warp factor.
As a result, without a brane the variable $x$ extends through the
full interval of $[-1,1]$. It follows from the theory of spherical
harmonics that normalizability restricts $l$ to positive integers,
$l\ge|m|$.  In this case, this indicates that $l=2,3,4,\ldots$, or
$E_0=4,5,6,\ldots$.

Inverting the spin-2 relation (\eqmathra) for the graviton mass, we find
the zero tension spectrum
$$
\mu^2=E_0(E_0-3)k^2,\qquad E_0=4,5,6,\ldots,
\numbereq\name{\eqector}
$$
a result obtained previously by Karch and Randall [\karch] and which
agrees with the embedding of AdS$_4$ in AdS$_5$.
The normalizable wave function reads
$$
\psi_{E_0}(x)=\sqrt{{(2E_0-3)(E_0-4)!\over2E_0!}}{\kappa^{3/2}\over
k}(1-x^2)P_{E_0-2}^{\,2}(x).
\numbereq\name{\eqwarped}
$$
In this case there is no massless four-dimensional graviton, which is
understood since there is in fact no brane present.

{\bf \noindent\the\secnumber.2 With a brane}
\par\noindent
With the introduction of a brane, we now work on the right side of the
brane, $\bar x<x<1$, and impose the Israel matching condition $d\psi/dx
\vert_{x=\bar x}=0$.  The requirement that the solution satisfies this
condition now renders the $l$ (or equivalently $E_0$) non-integers.
Nevertheless, we retain the standard AdS$_4$ expression (\eqector) for the
mass with a non-integer $E_0=l+2$. 

For non-integral $l$, it is more convenient to work with the
representation of the associated Legendre functions in terms of
hypergeometric functions.  In particular,
$$
P_{E_0-2}^{\,2}(x)={\Gamma(E_0+1)\over 8\,\Gamma(E_0-3)}(1-x^2)
\,{}_2F_1(4-E_0,E_0+1;3;{1-x\over2}).
\numbereq\name{\eqbrink}
$$
We wish to investigate the Kaluza-Klein spectrum in the nearly flat
brane limit, corresponding to $k\ll \kappa$, or $\bar x\approx
-1+k^2/2\kappa^2$.  Thus we need to examine the behavior of $\psi$
near $x=-1$.  Since this corresponds to the argument of ${}_2F_1$
approaching one, we invoke the transformation theory of hypergeometric
functions.  Some of the relevant properties are given in Appendix~A.

\noindent{\it{Low lying modes}}

The low lying states are those with $\mu \ll \kappa$, for which the
quantum number $E_0$ remains close to an integer.  By writing
$\epsilon=(1+x)/2$, which is small near the brane, the function
$\psi$ takes the asymptotic form
$$
\eqalign{
\psi&\simeq C\left\{-{4\over\pi}[1+E_0(E_0-3)\epsilon]\sin E_0\pi
+2\fft{\Gamma(E_0+1)}{\Gamma(E_0-3)}\epsilon^2
\left[\cos E_0\pi\vphantom{\fft{E_0}{(E_0)}}\right.\right.\cr
&\qquad\quad\left.\left.+\fft1\pi\left(-\fft{3E_0+1}{2(E_0-1)}
+2\gamma+2\psi(E_0) +\ln\epsilon\right)\sin E_0\pi\right]
+O(\epsilon^3)\right\}.}
\numbereq\name{\eqkuchma}
$$
The matching condition then yields for $E_0$
$$
E_0=n+3+(n+1)(n+2){{k^2}\over {4\kappa^2}},
\numbereq\name{\eqpaul}
$$
where $n$ is a non-negative integer.  Note that $n=0$ yields
$E_0=3+k^2/2\kappa^2$, which corresponds to the quasi-zero mode.
In particular, using (\eqector), we see that this mode
has mass $\mu^2=3k^4/2\kappa^2$, which agrees with what was found in the
previous section.  Except for the quasi-zero mode, the wave
function (\eqwarped) with integer $E_0$ remains a good approximation
in the bulk.

\noindent{\it{Higher modes}}

For the higher eigenvalues, the omitted terms in the expansion of the
hypergeometric function must to be restored, and we need to evaluate
the confluent limit of a hypergeometric function with a
fine-tuned small argument and large parameters. In terms
of $x=-1+{\zeta^2/2n^2}$ with $\zeta={\cal O}(1)$ and $n \gg 1$, we have
$\zeta=(\mu/\kappa)e^{\kappa y}$ and the wavefunction becomes
$$
\psi=\sqrt{\kappa^3k\over \mu^3}\zeta^2
[J_2(\zeta)\cos n\pi+N_2(\zeta)\sin n\pi],
\numbereq\name{\eqkriss}
$$
where $J_2(\zeta)$ and $N_2(\zeta)$ are the second Bessel and Newmann
functions, respectively. The derivation
of this formula appears in Appendix A. The quantization condition becomes
$$
\tan{n\pi}=-{J_1({{\mu}\over{\kappa}})
\over N_1({{\mu}\over{\kappa}})}.
\numbereq\name{\eqcoma}
$$
This limit reproduces the Randall-Sundrum model with a flat brane. 

\newsection Brane Bending and the Radion Mode.%
In the previous two sections, we have demonstrated that the spectrum of
transverse-traceless gravitational modes is composed of a
quasi-zero mode graviton (trapped on the brane) as well as a discrete
Kaluza-Klein tower.  The mode expansion may be given in terms of the
four-dimensional $E_0$ eigenvalue (equivalently the KK mass) and the
eigenfunctions along the $y$ direction, $\psi(y)$.  It is important to
realize that this analysis was performed under the requirement of two
boundary conditions.  Firstly, normalizability of $\psi(y)$ demands
that it vanishes sufficiently fast at the AdS$_5$ boundary, $y\to\infty$.
And, secondly, the Israel matching condition in the absence of matter
imposes the vanishing of the derivative at $y=0$ ({\it i.e.}~on the
brane), namely $\dot\psi(0)=0$.  Consequently, values of $E_0$ less than
two were immediately ruled out. This follows from the asymptotic expression of
$\psi(y)$ as $y\to 0$, since for $E_0<2$ it is no longer possible to satisfy
the Israel matching condition on (\eqbravo) without introducing the
companion solution, which is not normalizable.

However, one important exception must be made to the above boundary
conditions.  Namely, a solution which blows up would in fact be
acceptable, provided the non-normalizable piece can be removed by 
an appropriate residual gauge transformation, (\eqrbcv).  These gauge
transformations connect the behavior of $\psi(y)$ near the brane to
the one near the boundary of AdS$_5$; the wave function obtained in this
manner satisfies both boundary conditions at the expense of moving the
brane away from $y=0$ in the new coordinates.  Because of this, the
transformation is commonly denoted as brane bending [\garriga,\gkr].

Inserting the parameters (\eqriot) into the residual transformation
(\eqrbcv), we find a general parametrization for the gauge transformations
$$
\eqalign{
\delta a&=-{1\over k^2}(e^{\bar a}{\bar a}^{\prime}{\chi}^{\prime}
-2k^2{\chi}){\dot A}+{\bar a}^{\prime}{\phi},\cr
\delta b&=-{1\over k^2}(2e^{\bar a}{\chi}^{\prime\prime}+
e^{\bar a}{\bar a}^{\prime}{\chi}^{\prime}
-2k^2{\chi}){\dot A}-{\bar a}^{\prime}{\phi}+2{\phi}^\prime,\cr
\delta c&=-{1\over k^2}({2\over r}e^{\bar a}{\chi}^{\prime}
-2k^2{\chi}){\dot A}+{2\over r}{\phi},\cr}
\numbereq\name{\xxmollere}
$$
where $\chi$ and $\phi$ are both functions of $r$ only.  Here it is
apparent that $\phi$ generates four-dimensional gauge transformations on
the brane as in (\eqdiop), while $\chi$ is responsible for
transformations involving the
bulk.  For the latter, the $y$-profile is given simply by $\dot A$.  It
is this (and only this) form of non-normalizable behavior that may be
canceled through brane bending.  Since $\psi\sim \dot A$ solves the
eigenmode equation (\eqwilde) with $\mu^2=-2k^2$, extra care must be
taken in analyzing this case.  This provides the origin of the
Karch-Randall radion, which we examine here.

For $\mu^2=-2k^2$, corresponding to $E_0=2$, the mode equation
(\eqwilde) admits two linearly independent solutions, 
$$
\psi_1=\dot A,
\numbereq\name{\eqwave1}
$$
as indicated above, and 
$$
\psi_2=(\kappa-\dot A)^2,
\numbereq\name{\wave2}
$$ 
which can be verified by direct substitution.  The $\psi_1$ mode, which
does not vanish at the AdS$_5$ boundary, has the same $y$-dependence as 
the brane bending term of (\xxmollere).  On the other hand, $\psi_2$ is
normalizable, and cannot be removed by gauge transformations.  The
appropriate linear combination of $\psi_1$ and $\psi_2$ that solves
the Israel matching condition is given by
$$
\psi_{\rm radion}=\psi_1+\ft12(\kappa-\dot A(0))^{-1}\psi_2
=\dot A+\fft12\fft{(\kappa-\dot A)^2}{\kappa-\dot A(0)}.
\numbereq\name{\radiondef}
$$
We identify this as the radion mode.

Since the radion corresponds to $E_0=2$, the radial function $b(r)$
is given by solutions to (\eqvios) with $\mu^2=-2k^2$.  However, instead
of solving (\eqvios) directly, we note that there is already a natural
parametrization for $b(r)$, given by the gauge transformation
(\xxmollere) itself.  This suggests that we choose an ansatz for the
radion mode according to
$$
\eqalign{
a_{\rm radion}(r,y)&\equiv a_{BR}(r)\psi_{\rm radion}(y)
=-{1\over k^2}(e^{\bar a}{\bar a}^{\prime}{\chi}^{\prime}
-2k^2{\chi})\psi_{\rm radion},\cr
b_{\rm radion}(r,y)&\equiv b_{BR}(r)\psi_{\rm radion}(y)
=-{1\over k^2}(2e^{\bar a}{\chi}^{\prime\prime}+
e^{\bar a}{\bar a}^{\prime}{\chi}^{\prime}
-2k^2{\chi})\psi_{\rm radion},\cr
c_{\rm radion}(r,y)&\equiv c_{BR}(r)\psi_{\rm radion}(y)
=-{1\over k^2}({2\over r}e^{\bar a}{\chi}^{\prime}-
2k^2\chi)\psi_{\rm radion}.\cr}
\numbereq\name{\eqmollere1}
$$
The transverse-traceless condition requires $\chi$ to satisfy the
homogeneous equation,
$$
{\chi}^{\prime\prime}+\left({2\over r}+{\bar a}^{\prime}\right)
{\chi}^\prime-4k^2e^{-{\bar a}}{\chi}=0,
\numbereq\name{\eqatwood1}
$$
which is simply the scalar equation
$$
[-\Box+4k^2]\chi=0,
\numbereq
$$
in the AdS$_4$ background.  Since the equation for a massive scalar in
AdS$_4$ has the form $[-\Box+E_0(E_0-3)]\varphi=0$, this identifies the
radion (as well as the corresponding brane bending) mode as an
$E_0=4$ scalar [\karch].  Provided this scalar equation is solved, it is
easy to verify that $b_{BR}$ solves the spin-2 equation (\eqvios) 
with $\mu^2=-2k^2$.  This formally corresponds to a spin-2 value of $E_0=2$,
but, as indicated by our choice of $\chi$, more properly should be
realized as a spin-0 value of $E_0=4$. 

The homogeneous equation (\eqatwood1) can be transformed to
hypergeometric type by introducing the variable $\xi=-k^2r^2$
$$
\xi(1-\xi){{d^2{\chi}}\over {d{\xi^2}}}+
(\ft32-\ft52\xi){{d{\chi}}\over {d{\xi}}}+\chi=0.
\numbereq\name{\eqcedar}
$$
The solution can be written as
$$
{\chi}(r)=C_1 \,{}_2F_1(2, -\ft12; \ft32; -k^2r^2)+
{{C_2}\over r}\,{}_2F_1(\ft32, -1; \ft12; -k^2r^2),
\numbereq\name{\eqviote}
$$
with
$$
\eqalign{
{}_2F_1(2, -\ft12; \ft32; -k^2r^2)&=\ft34+\ft14(1+3k^2r^2)
{{\tan^{-1}kr}\over {kr}},\cr
{}_2F_1(\ft32, -1; \ft12; -k^2r^2)&=1+3k^2r^2.}
\numbereq\name{\eqpori}
$$
The combination that vanishes as $r \to \infty$ reads
$$
{\chi}(r)=C\left[
{{1+3k^2r^2}\over {3kr}}{\tan^{-1}{1\over {kr}}}-1\right],
\numbereq\name{\eqdefa1}
$$
Substituting (\eqdefa1) to (\eqmollere1), we find
$$
\eqalign{
a_{BR}(r)&={4\over 3}C\left[{1\over {kr}}{\tan^{-1}{1\over {kr}}}
-{1\over {1+k^2r^2}}\right], \cr
b_{BR}(r)&={4\over 3}C{1\over {k^2r^2}}
\left[-{1\over {kr}}{\tan^{-1}{1\over {kr}}}-{1\over {1+k^2r^2}}\right], \cr
c_{BR}(r)&={2\over 3}C{1\over {k^2r^2}}
\left[{1-k^2r^2\over {kr}}{\tan^{-1}{1\over {kr}}}
+1 \right]. \cr}
\numbereq\name{\eqmollere2}
$$

As will be seen in the following section, the radion mode is
sourced by the trace of the energy-momentum tensor on the brane.  In
particular, it will contribute to the static metric with a point source.
Asymptotic expansion of (\eqmollere2) indicates that $b_{\rm radion}(r,y)
={\cal O}({1/r^4})$ for large $r$, which dominates over the quasi-zero mode 
graviton contribution to $b(r,y)$. This presents a challenge to the
recovery of the Schwarzschild-AdS$_4$ metric on the brane. However,
performing a transformation on the brane to bring the metric into a
standard form, $c_{\rm radion}(r,0)=0$, we obtain
$$
\eqalign{
a_{\rm radion}(r, 0)&={{2C}\over 3(1+k^2r^2)}\biggl(
{{1+3k^2r^2}\over {kr}}{\tan^{-1}{1\over {kr}}}-3\biggr)\psi_{\rm
radion}(0), \cr
b_{\rm radion}(r,0)&={{2C}\over {3(1+k^2r^2)}}\biggl(
{{1-k^2r^2}\over {kr}}
{\tan^{-1}{1\over {kr}}}+1 \biggr)\psi_{\rm radion}(0), \cr}
\numbereq\name{\eqmollere3}
$$
which now becomes subleading relative to the quasi-zero mode graviton.
While it would be important to determine if the radion mode remains
subleading beyond the linearized approximation, this is outside the
scope of our present analysis.

\newsection Matter Sources on the Brane.%
Having discussed the general features of the braneworld linearized gravity
system in the transverse traceless gauge, we now return to the issue of
matter sources on the brane.  Recall from Eq.~(\eqvirile) that the
trace of the stress-energy tensor on the brane provides an obstruction
to imposing the traceless condition, $c=-\fft12(a+b)$.  In particular,
for a static mass point on the brane, where 
$T_{00}=M{\delta^3}(\vec r\,)$, the constraint, (\eqvirile), takes the form
$$
a+b+2c=-{8\pi G_5M\over3k^2}{\dot A}(y){\delta^3}(\vec r\,)+f(r),
\numbereq\name{\eqdesai}
$$
where $f(r)$ is an arbitrary function of $r$ and we restrict our
attention to $y \ge 0$. Although the source is located on the brane,
the delta function survives as we move into the bulk.

Although this effect presents an obstruction to working in
transverse-traceless gauge, the $y$-profile of (\eqdesai), given by
$\dot A$, was identified in the previous section as that related to
brane bending and the radion.  In particular, Eq.~(\eqdesai) indicates
that the mass (or trace of the stress tensor on the brane) provides a
source for the $\psi_1$ mode of the previous section.  If inducing a
non-trivial $\psi_1$ where the only effect of a source on the brane,
then there would be no additional physical consequences, as this is
purely gauge.  However, as indicated by (\radiondef), $\psi_1$ fails
to satisfy the Israel matching condition by itself, and necessarily
comes in through the radion combination.  This demonstrates that the
radion, in fact, is being sourced by matter on the brane.

To provide a heuristic accounting for the radion, consider writing the
metric functions as
$$
b(r,y) = \tilde b_{TT}(r,y) + b_{\rm radion}(r,y) = \tilde b_{TT}(r,y) +
b_{BR}(r) \psi_{\rm radion}(y)
\numbereq
$$
(and similarly for $a$ and $c$).  Here, $\tilde b_{TT}$ corresponds to
the $E_0>2$ solutions of section 3 (quasi-zero mode graviton) and
section 4 (KK gravitons) in the transverse-traceless gauge.  Similarly,
$b_{\rm radion}$ is given as in section 6, with the exception that it is
no longer a traceless mode, but is sourced by (\eqdesai).  As a slight
complication, $\psi_{\rm radion}$ contains both $\psi_1$ and $\psi_2$,
with only $\psi_1$ sourced by (\eqdesai), and the $\psi_2$ mode
remaining traceless.  For this reason, it is more convenient, in fact,
to split off $\psi_2$, and incorporate it into $\tilde b_{TT}$.  Thus we
write instead
$$
b(r,y) = b_{TT}(r,y) + b_{BR}(r) \psi_1(y)
\numbereq\name{\eqmollere}
$$
(and again similarly for $a$ and $c$), where $b_{TT}$, while remaining
transverse-traceless, now has an $E_0=2$ component as well as the
usual $E_0>2$ contributions.

By substituting Eq.~(\eqmollere) into (\eqdesai), we find that the
scalar radion mode $\chi$ of (\eqatwood1) now satisfies an inhomogeneous
equation with source
$$
{\chi}^{\prime\prime}+\left({2\over r}+{\bar a}^{\prime}\right)
{\chi}^\prime-4k^2e^{-{\bar a}}{\chi}=
{{4\pi G_5M}\over 3}{\delta^3}(\vec r\,).
\numbereq\name{\eqatwood}
$$
This essentially fixes the constant of (\eqdefa1) to be
$$
C=-\fft{2G_{5}Mk}\pi.
\numbereq\name{\eqdefa}
$$
Consequently, the complete normalized radion mode $b_{\rm radion}$ is
expected to be
$$
\eqalign{
b_{\rm radion}(r,y)&=b_{BR}(r)\psi_{\rm radion}(y)\cr
&={{8G_{5}M}\over 3\pi}{1\over {kr^2}}
\biggl({1\over {kr}}{\tan^{-1}{1\over {kr}}}+{1\over {1+k^2r^2}}\biggr)
\biggl(\dot A+\fft12{{(\kappa-{\dot A})^2}\over {\kappa-{\dot A}(0)}}
\biggr),}
\numbereq\name{\eqstravinsky}
$$
with the $\psi_1$ term arising from $\chi$ and the $\psi_2$ term arising
from $b_{TT}$.  Only the former may be removed by a brane bending gauge
transformation, leaving the component proportional to $\psi_2$ as the
physical radion mode. 

The determination of $a_{TT}$, $b_{TT}$ and $c_{TT}$ of the solution 
(\eqmollere) involves the inversion of the Lichnerowicz operator with a 
AdS$_4$ background, which can be made explicit for this special case. 
As is always the case for an elliptic equation of two variables, the
solution can be written either as a sum over the modes (obtained
in the previous sections) which are oscillatory in $y$ or as a sum
over modes which are oscillatory in $r$. The former approach highlights
the asymptotic AdS$_4$ behavior on the brane for $r\to\infty$ but
overshadows the damping behavior off the brane, especially for 
higher modes.  For a source only on the brane, the off-brane damping 
(which is implicit in the summation over the KK modes) is an important
feature of the solution.  To make this more explicit, we now switch to
the latter (oscillatory in $r$) approach in constructing
integral representations of $a_{TT}$, $b_{TT}$ and $c_{TT}$.

On writing $\mu^2/k^2=-{9\over 4}-\nu^2$, which corresponds to complex
$E_0={3\over 2}+i{\nu}$, equation (\eqvios) defines
a Sturm-Liouville problem with a continuous spectrum $\nu^2$
and measure $r^4(1+k^2r^2)$ for the normalization
integral. The radial eigenfunctions read
$$
b_\nu(r)\equiv b(r|\mu)\big\vert_{\mu^2/k^2=-9/4-\nu^2}
={}_2F_1(\ft74+\ft{i}2\nu,\ft74-\ft{i}2\nu; \ft52; -k^2r^2),
\numbereq\name{\eqtomlins}
$$
and display the asymptotic behavior
$$
b_\nu(r) \simeq {\sqrt 2}C_{\nu}^{-1}{{\sin(\nu\,{\ln kr}+{\delta}_{\nu})}
\over {(kr)^{7\over 2}}},
\numbereq\name{\eqwhitman}
$$
for large $kr$. The phase shift $\delta_\nu$ is given by
$$
e^{2i{\delta}_\nu}=-{{{\Gamma(i\nu)}{\Gamma({3\over 4}-\fft{i}2\nu)}
{\Gamma({7\over 4}-\fft{i}2\nu)}}\over {{\Gamma(-i\nu)}
{\Gamma({3\over 4}+\fft{i}2\nu)}
{\Gamma({7\over 4}+\fft{i}2\nu)}}},
\numbereq\name{\eqwalt}
$$
while the normalization constant is
$$
C_{\nu}^2={2\over {9\pi^2}}({\nu}^2+\ft94){\Gamma^2(\ft34
+\ft{i}2\nu)}{\Gamma^2(\ft34-\ft{i}2\nu)}{\nu}\,
{\sinh{\pi\nu}}.
\numbereq\name{\eqcaros}
$$
The corresponding solution of (\eqwilde) that vanishes on the AdS$_5$
boundary is
$$
\eqalign{
\psi_\nu(y)&\equiv\psi(y|\mu)\big\vert_{\mu^2/k^2=-9/4-\nu^2}\cr
&=\sech^4{\kappa}(y-y_0)
\,{}_2F_1(\ft52+i{\nu}, \ft52-i{\nu};3; {{1-{\tanh{\kappa(y-y_0)}}}\over
2}).}
\numbereq\name{\eqcrap}
$$
This allows us to decompose the transverse-traceless function $b_{TT}$
in terms of a spectral density $\beta_\nu$ as
$$
\eqalign{
b_{TT}(r,y)&=\int_0^\infty\fft{d\nu}\pi\beta_\nu\,
b_\nu(r)\fft{\psi_\nu(y)}{\dot\psi_\nu(0)}\cr
&=\int_0^\infty\fft{d\nu}\pi\beta_\nu\,
{}_2F_1(\ft74+\ft{i}2\nu,\ft74-\ft{i}2\nu,\ft52;-k^2r^2)
\fft{\psi_\nu(y)}{\dot\psi_\nu(0)}.}
\numbereq\name{\eqerrikos}
$$

In order to determine $\beta_\nu$, we apply the Israel matching
conditions, $\dot b(r,y)=0$ to (\eqmollere).  The result is simply
$\dot b_{TT}(r,0)=-b_{BR}(r)\ddot A(0)$, which reads more explicitly
$$
\int^\infty_0{{d\nu}\over {\pi}}
{\beta}_{\nu}\,
{}_2F_1(\ft74+\ft{i}2\nu, \ft74-\ft{i}2\nu;\ft52; -k^2r^2)
=-{8G_5M\over 3\pi}\fft1{kr^2}
\left({1\over kr}{\tan^{-1}{1\over kr}}+ {1\over {1+k^2r^2}} \right)
{\ddot A}(0).
\numbereq\name{\eqinvert}
$$
Notice that the $\delta^{3}(\vec r\,)$ function pertaining to the
source is weak with respect to the measure $r^4(1+k^2r^2)$ and
subsequently is omitted from the matching condition.  We now invert this
integral to solve for $\beta_\nu$.  To do so, we use the result of
Appendix B, which demonstrates that any square integrable function of
$r$ with respect to the measure $r^4(1+k^2r^2)$ can be expressed as a
generalized Fourier integral
$$
f(r)={\int^{\infty}_{0}}{{d\nu}\over {\pi}}{\cal F}(\nu)\,
{}_2F_1(\ft74+\ft{i}2\nu,\ft74-\ft{i}2\nu;\ft52; -k^2r^2),
\numbereq\name{\eqionesco}
$$
with $\cal F$ given by the inverse transformation
$$
{\cal F}(\nu)=C_{\nu}^2k^5{\int^{\infty}_{0}}dr\,r^4(1+k^2r^2)f(r)\,
{}_2F_1(\ft74+\ft{i}2\nu,\ft74-\ft{i}2\nu; \ft52; -k^2r^2).
\numbereq\name{\eqenescu}
$$
It thus follows that
$$
{\beta}_{\nu}=-{{4G_{5}Mk}\over 3\pi}{\ddot A}(0)
C_{\nu}^2{\int^{\infty}_0}dx\,
\left[(1+x){\tan^{-1}{1\over {\sqrt x}}}+{\sqrt x}\right]\,
{}_2F_1(\ft74+\ft{i}2\nu, \ft74-\ft{i}2\nu;\ft52; -x).
\numbereq\name{\eqsferis}
$$
Using the identity
\ref{\bat}{A. Erdelyi, {\it et al.}, {\it Higher Transcendental
Functions}, vol.1 (McGraw-Hill, 1953).}
$$
(1+x)\,{}_2F_1(\ft74+\ft{i}2\nu, \ft74-\ft{i}2\nu;
\ft52; -x)=-{6\over {{\nu}^2+{1\over 4}}}
{d\over {dx}}\Bigl[(1+x)^2\,{}_2F_1(\ft74+\ft{i}2\nu, 
\ft74-\ft{i}2\nu;\ft32; -x)\Bigr],
\numbereq\name{\eqhyp}
$$
and integrating by parts, we find that
$$
\eqalign{
{\beta}_{\nu}=-4G_5Mk{\ddot A}(0){C_{\nu}^2\over {\nu}^2+\fft14}
-\fft{4G_5Mk}{3\pi}\ddot A(0)C_\nu^2&\int_0^\infty\fft{dx}{\sqrt{x}}
\biggl[x\,{}_2F_1(\ft74+\ft{i}2\nu,\ft74-\ft{i}2\nu;\ft52;-x)\cr
&-3\fft{1+x}{\nu^2+\fft14}\,
{}_2F_1(\ft74+\ft{i}2\nu,\ft74-\ft{i}2\nu;\ft32;-x)\biggr].}
\numbereq\name{\eqritsos}
$$
Using formulae for the integration of a hypergeometric function
times a power function, we find that the integral on the right
hand side of (\eqritsos) vanishes, and we are left with a simple
expression for $\beta_\nu$.  As a result, $b_{TT}(r, y)$ reads
$$
b_{TT}(r, y)=-4G_{5}Mk{\ddot A}(0)
{\int^{\infty}_{0}}\fft{d\nu}\pi{{C_{\nu}^2}\over {{\nu}^2+{1\over
4}}}\,
{}_2F_1(\ft74+\ft{i}2\nu, \ft74-\ft{i}2\nu;\ft52; -k^2r^2)
\fft{\psi_\nu(y)}{\dot\psi_\nu(0)}.
\numbereq\name{\eqferil}
$$
The corresponding expressions for $a_{TT}$ and $c_{TT}$ follow from
the traceless and transverse conditions.  Then, by substituting the
expressions back into (\eqmollere), we obtain the explicit solution to
the linearized Einstein equations.

To relate the integral representation found above to the summation
over Kaluza-Klein modes, we convert (\eqferil) into a contour integral
$$
\eqalign{
b_{TT}(r, y)={{8G_{5}Mk}\over 3\sqrt{\pi}}{\ddot A}(0)
&{\int^{\fft32+i\infty}_{\fft32-i\infty}}{{dE_0}\over {2{\pi}i}}
{E_0\Gamma^2(E_0/2)\over
(E_0-1)(E_0-2)\Gamma(E_0-\fft32)}(kr)^{-E_0-2}\cr
&\times\,{}_2F_1(1+\ft12E_0,-\ft12+\ft12E_0;-\ft12+E_0;-(kr)^{-2})
\fft{\psi_{i(E_0-\fft32)}(y)}{\dot\psi_{i(E_0-\fft32)}(0)},\cr}
\numbereq\name{\eqril}
$$
where we have used the expression for $C_\nu^2$ given in Eq.~(\eqcaros).
To highlight the $r\to\infty$ behavior of $b_{TT}$, we have in addition
used the relation between hypergeometric functions of arguments $\zeta$
and ${1/{\zeta}}$.  Enclosing the contour from the right half plane,
we pick up the residues of the poles along the positive real axis (see
Fig.~1). In addition to the trapped graviton and Kaluza-Klein modes
(previously identified in Sections 3 and 4) which are specified by the
condition $\dot\psi_{i(E_0-\fft32)}(0)=0$ (which is simply the Israel
matching condition), we find a pole at $E_0=2$ corresponding to the
presence of a radion mode; its residue produces exactly the term of
(\eqstravinsky) proportional to $b_{BR}(r)\psi_2(y)$.

\noindent{\it Asymptotic expressions on the brane}

To get a better understanding of the integral representation for
$b_{TT}$, we turn to several asymptotic limits.  We first explore the 
Randall-Sundrum limit by taking $k\to 0$ [or equivalently
$y_0\to\infty$, as seen from (\kkyrel)], while keeping
$\kappa y ={\cal O}(1)$.  Introducing $p=\nu k$ and $\zeta={e^{\kappa y}/
\kappa}$, we note that
$$
\tanh\kappa(y-y_0)\simeq -1+{k^2\zeta^2\over 2}.
\numbereq\name{\RSlimit}
$$ 
It follows that the behavior of the hypergeometric representation
(\eqcrap) of $\psi_\nu(y)$ singles out $p\zeta={\cal O}(1)$.
Highlighting this behavior motivates us to split the range of
integration of Eq.~(\eqferil) according to
$$
\int_0^\infty\fft{d\nu}\pi=
\int_0^{\nu_0}\fft{d\nu}\pi+
\int_{\nu_0}^\infty\fft{d\nu}\pi=
\int_0^{\nu_0}\fft{d\nu}\pi+{1\over k}\int_{k\nu_0}^\infty\fft{dp}\pi,
$$
where $\nu_0$ is fixed, and satisfies $1\ll\nu_0\ll\kappa/k$.
Correspondingly, we write
$$
b_{TT}(r,y)= b_{TT}^{LL}(r,y)+ b_{TT}^{RS}(r,y).
\numbereq\name{\dividex}
$$
We now treat $b_{TT}^{RS}$ and $b_{TT}^{LL}$ separately, according
to the range of integration.

To estimate the contribution of $b_{TT}^{LL}$, we approximate
the hypergeometric function in $kr\ll1$ of the integrand by unity and 
$$
\psi_\nu(y)\simeq
{{32}\over {{\Gamma}({5\over 2}+i{\nu})
{\Gamma}({5\over 2}-i{\nu})}}
[1-\ft14(\nu^2+\ft94)k^2\zeta^2].
\numbereq\name{\approxy}
$$
These Gamma functions near the upper limit may be estimated with the Stirling 
formula. The result is that
$$
b_{TT}^{LL}(r,y)\simeq{16\over 3\pi}{\ddot A}(0)\kappa\nu_0.
\numbereq\name{\transition}
$$
The cutoff $\nu_0$ can be sent to zero in the limit $k\to 0$. 

For $b_{TT}^{RS}$, on the other hand, the hypergeometric functions in the 
integrand are dominated by their confluent limits.  Following the steps
outlined in Appendix A, we find that
$$
{}_2F_1(\ft74+\ft{i}2\nu,\ft74-\ft{i}2\nu;\ft52;-k^2r^2)
\simeq {3\over pr}j_1(pr),
\numbereq\name{\sphere}
$$
while
$$
\psi_\nu(y)
\simeq {16\over\pi\nu^2}\zeta^2K_2(p\,\zeta)\cosh\pi\nu.
\numbereq\name{\modify}
$$
Combining these expressions, we find that in the Randall-Sundrum limit 
$b_{TT}(r,y)$ becomes
$$
b_{TT}^{RS}(r,y)={{8G_5M}\over 3\pi}
e^{2\kappa y}\int_0^\infty dp\,p{j_1(pr)\over pr}
{K_2(p\,\zeta)\over K_1({p/\kappa})},
\numbereq\name{\RSresult}
$$
which agrees with the expression obtained previously in Ref.~[\ighcr].

In the region where $1/k \gg r \gg 1/\kappa$ on the brane, the integral
(\RSresult), for $b_{TT}^{RS}$ may be approximated.  The result is
$$
b_{TT}^{RS}(r, 0) \simeq {{4G_4M}\over 3r}
+{{4G_4M}\over {3{\kappa}^2r^3}}({\ln{2{\kappa}r}}-1),
\numbereq\name{\eqcries}
$$
where we have used the two-sided Randall-Sundrum relation $G_4=\kappa
G_5$.  We must take one more step before examining the geometry, and
that is to account for the $b_{BR}$ mode, which in the $k\to0$
limit reads $b_{BR}(r) \simeq 4G_5M/3k^2 r^3$.  Substituting $b_{TT}$ and
$b_{BR}$ into (\eqmollere), and using $\psi_1(0)\simeq -\kappa$ as well
as the transverse-traceless conditions to recover $a_{TT}$ and $c_{TT}$,
we find
$$
\eqalign{
a & \simeq -{{4G_4M}\over 3r}\left(1+\fft1{\kappa^2r^2}\right),\cr
b & \simeq {{4G_4M}\over 3r}\left(1-\fft1{k^2r^2}+\fft1{\kappa^2r^2}
(\ln{2\kappa r}-1)\right),\cr
c & \simeq {{2G_4M}\over {3r}}\left(\fft1{k^2r^2}-\fft1{\kappa^2r^2}
(\ln{2\kappa r}-2)\right).}
\numbereq\name{\eqpios}
$$
By performing a gauge transformation to set $c=0$, the metric on the brane
is finally transformed into the standard form
$$
ds^2=-\left[1-\fft{2G_4M}r\left(1+{2L_5^2\over {3r^2}}\right)\right]dt^2
+\left[1+\fft{2G_4M}r\left(1+{L_5^2\over{r^2}}\right)\right]dr^2
+r^2d\Omega^2,
\numbereq\name{\eqcallos}
$$
which is simply the linearized Schwarzschild solution including the
first Randall-Sundrum correction [\garriga,\duffliu].  This reproduces
the Newtonian potential, (\newtcor).

We now return to the Karch-Randall case, and examine the large distance
behavior, $kr\gg1$.  The large $r$ behavior of $b_{TT}(r,y)$ is easily
obtained through its contour integral representation, (\eqril),
which enables us to write
$$
b_{TT}(r,y)=\bar b_{\rm radion}(r,y)+b_0(r,y)+\sum b_{KK}(r,y).
\numbereq\name{\expansion}
$$
This is the decomposition as a sum over residues of the poles in
(\eqril).  Here, $\bar b_{\rm radion}(r,y)\sim b_{BR}(r)\psi_2(y)$ is
the $E_0=2$ residue corresponding to the physical radion,
$b_0(r,y)$ is the $E_0\approx3$ quasi-zero mode discussed in Section 3, 
and $b_{KK}$ represents the $E_0\approx4,5,6,\ldots$ Kaluza-Klein modes
discussed in Section 4. 

For large $kr$, the quasi-zero mode takes the form
$$
b_{0}(r, 0) \simeq -{{4G_5M{\kappa}}\over {3k}}{\ddot A}(0)
(kr)^{-5-{k^2/{2{\kappa}^2}}},
\numbereq\name{\eqjill}
$$
while the sum over the KK modes is dominated 
by its first term, which corresponds to the pole at 
$E_0\simeq 4+{3k^2/2\kappa^2}$.  Retaining only the leading term, we have
$$
\sum b_{KK}(r,0)\simeq-{16G_5Mk\over9\pi\kappa}
{\ddot A}(0)
(kr)^{-6-{3k^2/ 2\kappa^2}}+\cdots.
\numbereq\name{\eqKKmode}
$$
In general, it is clear from (\eqril) that the large $kr$ behavior
of the residue at the pole for $E_0$ is simply $b_{E_0}(r,0)\sim
(kr)^{-2-E_0}$.  So in fact the large $kr$ behavior of $b(r,0)$ itself
is dominated by that of the radion mode
$$
b_{\rm radion}(r,0) \simeq {{4G_5Mk}\over{3\pi\kappa}}\ddot
A(0)[(kr)^{-4}-\ft23(kr)^{-6}+\cdots].
\numbereq\name{\eqcaress}
$$
Combining the above expressions, this gives for $b(r,0)$
$$
b(r, 0)\simeq\fft{4G_4Mk}3\left[
\fft1\pi\left(\fft{k}\kappa\right)^2(kr)^{-4}
-(kr)^{-5-k^2/2\kappa^2}-\fft4{3\pi}\left(\fft{k}\kappa\right)^2
(kr)^{-6-3k^2/2\kappa^2}+\cdots\right].
\numbereq\name{\eqfraternity}
$$
Finally, we recover $a(r,0)$ and $c(r,0)$ from the transverse-traceless
conditions and perform a gauge transformation to set $c(r,0)=0$.  This
yields the linearized Karch-Randall solution
$$
\eqalign{
a & \simeq -2G_4Mk\left[(kr)^{-3-k^2/2\kappa^2}+\fft8{3\pi}
\left(\fft{k}\kappa\right)^2(kr)^{-4-3k^2/2\kappa^2}+\cdots\right],\cr
b & \simeq 2G_4Mk\left[(kr)^{-3-k^2/2\kappa^2}
-\fft2{9\pi}\left(\fft{k}\kappa\right)^2 (kr)^{-4}+\fft{32}{9\pi}
\left(\fft{k}\kappa\right)^2(kr)^{-4-3k^2/2\kappa^2}+\cdots\right].}
\numbereq\name{\eqcretinos}
$$
This series representation generalizes the metric obtained from the
quasi-zero mode, (\eqpoilas), to include the contributions from the
radion and KK modes.  However, unlike in the Randall-Sundrum case,
(\eqcallos), where only every other power of $r$ enters the metric, here
all integer powers plus fractional mass shifts contribute.

We note that the radion first enters $a$ at ${\cal O}(r^{-6})$,
and even in $b$ is no longer dominant over the quasi-zero mode graviton.  
Nevertheless, the radion does contribute at subleading order, and its
omission in $b$ would lead to an incorrect prediction of the
Karch-Randall corrections to Schwarzschild-AdS.  Of course, this is
mainly of academic interest, as the expansion of (\eqcretinos) is only
valid in the $kr\gg1$ regime.  Furthermore, it is important to realize
that the presence of the radion does not lead to a VVZ discontinuity, as
the conventional Randall-Sundrum metric, (\eqcallos), is obtained
smoothly in the $k\to 0$ limit.

\newsection Conclusions.%
In this paper, we have investigated the metric fluctuations of the
Karch-Randall model in a linearized gravity framework.  While our approach
has mainly focused on the properties of a static, axially symmetric metric
in $D=4+1$ dimensions giving rise to an AdS$_4$ braneworld, much of the
analysis of the Kaluza-Klein modes is independent of the metric on the
brane, and is hence completely general (at least within the limitations
of linearized gravity).  In first examining the source-free (on the
brane) Einstein equations, we presented an elementary procedure for
obtaining the mass of the (quasi-zero mode) Karch-Randall graviton.
Despite its mass, we demonstrate that the presence of this massive graviton
continues to reproduce the Schwarzschild-AdS$_4$ metric at large
distances on the brane.

Of course, the notion of mass in anti-de Sitter 
space is rather different from that in asymptotic Minkowski space.  In
AdS, the large distance behavior, $r\gg L$ (or, equivalently, $kr\gg1$)
is governed primarily by the AdS curvature.  Thus wavefunctions
naturally fall off extremely rapidly at large distances, regardless of
mass.  In the reference AdS$_4$ background, (\eqglobads), the falloff is
essentially given by $r^{-E_0}$, as may be seen in the power behavior of
(\eqcretinos).  A more directly observable effect of mass ({\it i.e.}~Yukawa
behavior) presumably may be seen only within an AdS radius.
However, for a Karch-Randall graviton of mass (\krmasse), with Compton
wavelength quadratic in $L$, this mass is essentially vanishing, and
presumably there is no regime where Yukawa behavior may be observed.
For this same reason, we believe a massive graviton as given here does
not signify a new screened phase of gravity, regardless of how general
covariance is broken (or whether it is even broken, as in the dynamical 
mass generation mechanism of Refs.~[\porhiggs,\duffliutwo]).  This
protection against screening is safe in the Minkowski limit, as any
finite $E_0$ leads to vanishing mass when the cosmological constant is
removed.

A somewhat unexpected result of this linearized gravity analysis is the
discovery of a physical radion in the (one brane) Karch-Randall model.
The radion is an $E_0=4$ scalar mode coupling to the trace of the
stress-energy tensor on the brane, and gives rise to a physical
correction to the Schwarzschild-AdS metric.  In the present case,
the radion arises through an incomplete cancellation of the $E_0=4$
scalar fluctuation with a similar brane-bending mode (and would be absent
in the RSII model, where the cancellation is complete).  While the
radion is directly sourced by matter on the brane, we have demonstrated
that it is always present in the spectrum, even in the absence of
sources.  This observation agrees with the result of [\chacko], which
demonstrated that the radion in a two-AdS-brane scenario survives in
the limit when the second brane is removed.
Although this analysis was only performed for static
configurations (and in addition with axial symmetry), the existence
of the radion mode may be traced to solutions of the KK mode
equation (\eqwilde), and hence is insensitive to any particular
metric ansatz on the brane.  This suggests that the radion survives as a
completely dynamical mode on the brane, and not simply as a constrained
trace mode of the graviton.

One may worry that the large distance expansion of $b(r)$ in (\eqcretinos)
is suggestive of a ghost, as the radion enters with the ``wrong'' sign.
However, this is merely an illusion, as the radion correction to the
Newtonian potential, hidden in the subleading behavior of $a(r)$, enters
physically.  This agrees with the results of
\lref{\papa}{A. Papazoglou, 
{\sl Dilaton Tadpoles and Mass in Warped Models},
\pl505 231 (2001) [arXiv:hep-th/0102015].
}%
[\papa,\chacko]
where it was demonstrated that the model is perturbatively stable.

In order to explain the physical origin of the Karch-Randall radion, we
note that unlike in the Randall-Sundrum model, where the position of the
brane may be removed by a conformal rescaling (and hence is unphysical),
for the AdS braneworld, the distance $y_0$ from the brane to the minimum of
the double exponential warp factor is not easily removed.  Only in the
Randall-Sundrum limit, when $y_0$ is pushed to infinity, does this
special point disappear.  It is the promotion of this $y_0$ distance to
a dynamical field that gives rise to a radion.  In this way, this
feature is similar to the RSI model, where the distance between the two
branes is governed by a radion (which similarly disappears when the
second brane is removed to infinity).  Of course, one may argue that
there is no preferred point such as $y_0$ in the bulk AdS$_5$ space.
While this is certainly the case, it is important to keep in mind that
the braneworld is not given simply by AdS$_5$, but depends crucially on how
it is sliced up.  After all, it is this difference that gives rise to
distinct classes of dS, flat and AdS braneworlds, which are all embedded
in AdS$_5$.  An important aspect of this slicing of AdS would be the
identification of the radion in a holographic interpretation of the
Karch-Randall model.

\vskip .1in
\noindent
{\bf Acknowledgments.} \vskip .01in \noindent
We would like to thank M.~Porrati for useful discussions.  JTL wishes to
thank D.~Gross and E.~Witten for discussions related to the interpretation
of mass in anti-de Sitter space. We would also like to thank
A.~Nelson and A.~Papazoglou for bringing to our attention
Refs.~[\chacko], [\defa] and [\papa].
This work was supported in part by the Department of Energy under Grants
Number DE-FG02-91ER40651-TASK~B and DE-FG02-95ER40899-TASK~G.

\newappendix Appendix A.%
In section 4, we made use of several analytic properties of
hypergeometric functions.  Here we summarize some of the
technical issues that were involved.  Firstly, for the off-brane
wavefunction (\eqbrink), the analytical continuation of
${}_2F_1(\alpha, 5-\alpha; 3; \zeta)$ to the neighborhood of $\zeta=1$ is
given by [\bat]
$$
\eqalign{
{}_2F_1(\alpha, 5-\alpha; 3; \zeta)&={2\over {{\Gamma(\alpha)}
{\Gamma(5-\alpha)}}} \Big [ {1\over {(1-\zeta)^2}}+
{{(2-\alpha)(3-\alpha)}\over {(1-\zeta)}} \Big ]\cr
&+{2\over {{\Gamma(3-\alpha)}
{\Gamma(-2+\alpha)}}}{\sum^{\infty}_{l=0}}
{{(\alpha)_l(5-\alpha)_l}\over {l!(l+2)!}}\Big [
{\psi}(l+1)+{\psi}(l+3)\cr
&-{\psi}(\alpha+l)-{\psi}(5-\alpha+l)
-{\ln{(1-\zeta)}}\Big ](1-\zeta)^{l}.\cr}
\numbereq\name{\eqkariaiva}
$$
Using the relation
$$
{\Gamma}(\zeta){\Gamma}(1-\zeta)={{\pi}\over {{\sin{\pi\zeta}}}},
\numbereq\name{\eqvaca}
$$
and its logarithmic derivative
$$
{\psi}(\zeta)-{\psi}(1-\zeta)=-{\pi}{\cot{\pi\zeta}},
\numbereq\name{\eqmaca}
$$
Eq.~(\eqkariaiva) can be written as
$$
{}_2F_1(-n, 5+n; 3; \zeta)={}_2F_1(-n, 5+n; 3; 1-\zeta){\cos{n\pi}}
+G(-n, 5+n; 3; 1-\zeta){\sin{n\pi}},
\numbereq\name{\eqkinos}
$$
where
$$
\eqalign{
G(-n, 5+n; 3;&1-\zeta)=\cr
&-{2\over {\pi}}
\Big [ {1\over {(n+1)(n+2)(n+3)(n+4)(1-\zeta)^2}}+
{1\over {(n+1)(n+2)(1-\zeta)}} \Big ]\cr
&-{2\over {\pi}}{\sum^{\infty}_{l=0}}
{{(-n)_l(5+n)_l}\over {l!(l+2)!}}\Big [
{\psi}(l+1)+{\psi}(l+3)\cr
&\kern2cm-{\psi}(n-l+1)-{\psi}(5+n+l)
-{\ln{(1-\zeta)}}\Big ](1-\zeta)^{l}.\cr}
\numbereq\name{\eqmanolis}
$$
The order by order approximation for small $1-\zeta$ at fixed $n$ can
be obtained readily, thus giving rise to Eq.~(\eqkuchma).

Secondly, to obtain (\eqkriss), we consider the series representation of
${}_2F_1(-n, 5+n; 3; 1-\zeta)$ with $1-\zeta
={{\xi}^2/{4n^2}}$:
$$
{}_2F_1(-n, 5+n; 3; {{\xi}^2\over {4n^2}})={\sum_{l=0}^{\infty}}
{{(-n)_l(5+n)_l}\over {l!(l+2)!}}\Big ({{\xi}\over {2n}} \Big )^{2l}.
\numbereq\name{\eqrotha}
$$
As $n \to \infty$, the contribution of the successive terms
drops quickly before $l$ becomes comparable to $n$. Therefore
the leading order approximation amounts to setting $(-n)_l \simeq
(-n)^l$ and $(5+n)_l \simeq n^l$. The series (\eqrotha) then
becomes proportional to that of the Bessel function $J_2$:
$$
{}_2F_1(-n, 5+n; 3; {{\xi}^2\over {4n^2}}) \simeq 2({2\over {\xi}})^2
J_2({\xi}).
\numbereq\name{\eqmaria}
$$
Using the same approximation, we may set in addition
$\psi(n-l+1)\simeq{\psi}(5+n+l)\simeq{\ln n}$.  Hence the
series representation of $G(-n, 5+n; 3; {{\xi}^2\over {4n^2}})$
becomes
$$
G(-n, 5+n; 3; {{\xi}^2\over {4n^2}}) \simeq 2({2\over {\xi}})^2
N_2({\xi}).
\numbereq\name{\eqmariaki}
$$
The wavefunction for the higher modes, (\eqkriss), is easily obtained by
substituting (\eqmaria) and (\eqmariaki) into (\eqkinos).  Finally,
working out the confluent limit of the function $b(r|\mu)$ to arrive at
(\sphere) follows in a similar manner.

\newappendix Appendix B.%
In this appendix, we derive the generalized Fourier transformation
utilized in section 6.  For proper normalization, we begin by
introducing a large box with
$$
0 < r < {1\over k}e^{L}.
\numbereq\name{\eqlios}
$$
Furthermore, we impose Dirichlet boundary conditions on function
$u(r|\mu)$, {\it i.e.}
$$
u({1\over k}e^{L}|\mu)=0.
\numbereq\name{\eqpios}
$$
The spectrum $\nu^2$ becomes discrete and is determined by
the condition
$$
{\nu}L+{\delta}_{\nu}=n{\pi}, \qquad n=0, 1, 2, \ldots,
\numbereq\name{\eqsios}
$$
for ${\nu}L \gg 1$. Using the asymptotic formula, we find
the normalization integral
$$
{\int^{{1\over k}e^{L}}_{0}}drr^4(1+k^2r^2)u^2(r|\mu)=
{L\over {C_{\nu}^2k^5}}.
\numbereq\name{\eqtios}
$$
The normalized eigenfunction now reads
$$
{\hat u}(r|\mu)={{C_{\nu}k^{5\over 2}}\over {\sqrt L}}
{}_2F_1(\ft74+\ft{i}2\nu,\ft74-\ft{i}2\nu;\ft52; -k^2r^2).
\numbereq\name{\eqgios}
$$
It follows from Sturm-Liouville theory that ${\hat{u}}(r|\mu)$ forms a
complete basis set according to which any function can be expanded:
$$
f(r)={1\over {\sqrt L}}{\sum_{\nu}}B_{\nu}{\hat u}(r|\mu),
\numbereq\name{\eqfios}
$$
and
$$
B_{\nu}={\sqrt L}{\int^{\infty}_{0}}dr\,r^4(1+k^2r^2){\hat u}(r|\mu)f(r).
\numbereq\name{\eqwios}
$$
In the limit $L\to\infty$, ${\sum_{\nu}}\to {L\over {\pi}}
\int^{\infty}_{0}d{\nu}$, and Eqs.~(\eqfios) and (\eqwios) become the
generalized Fourier transformation, (\eqionesco) and (\eqenescu), of
section 5.

\immediate\closeout1
\bigbreak\bigskip

\line{\twelvebf References. \hfil}
\nobreak\medskip\vskip\parskip

\input refs


\vfill\break
\line{\twelvebf Figures. \hfil}
\nobreak\medskip\vskip\parskip

\bigskip
\bigskip
\epsfxsize 3truein
\centerline{\epsffile{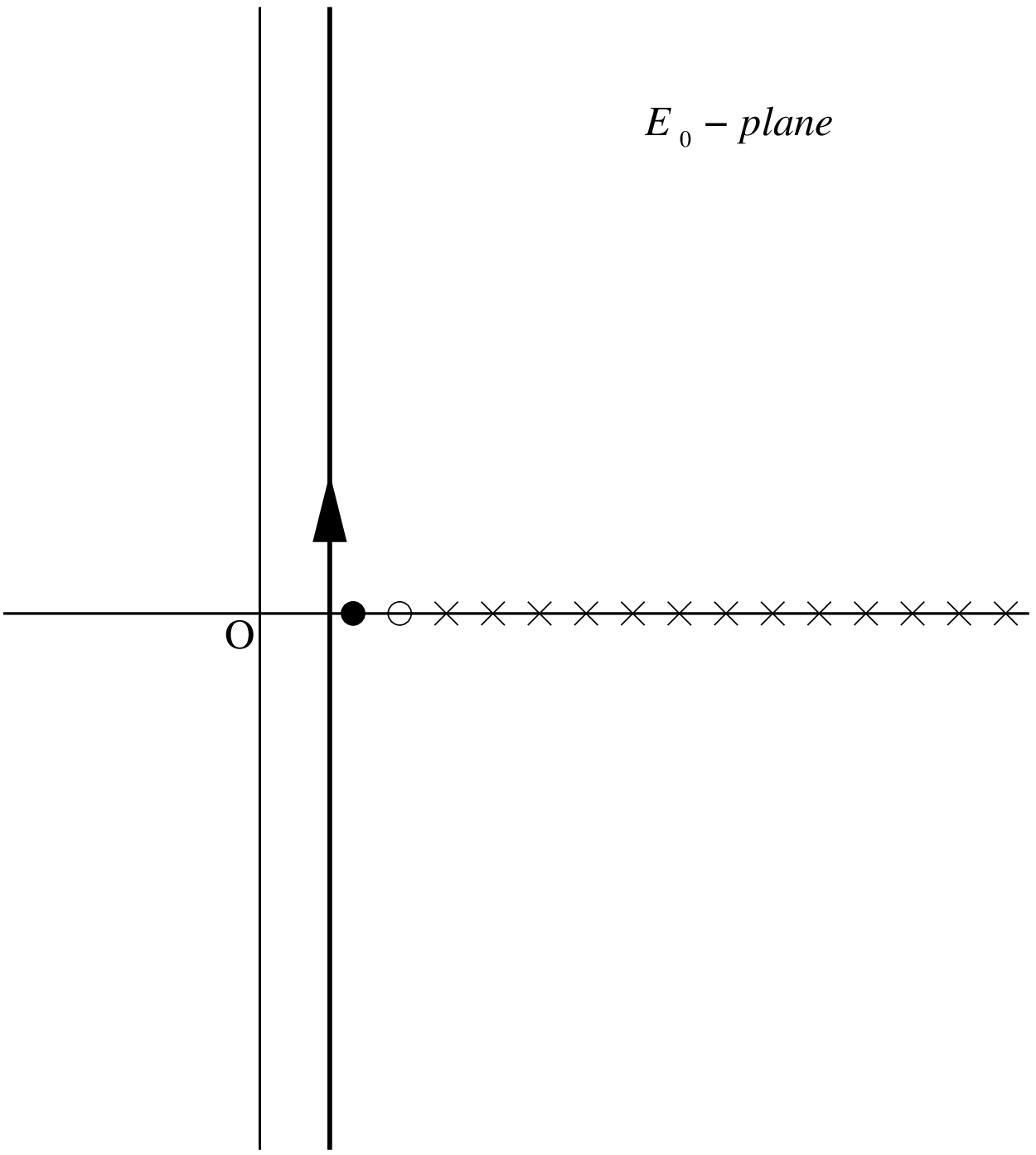}}
\medskip
{\narrower
\noindent
{\bf Figure 1:}
The integration contour on the complex $E_0$-plane and the poles
to the right of the contour. The solid circle denotes the radion
mode, the open circle represents the quasi-zero mode and the
crosses the KK modes.\par
}

\vfil\end

\bye